\documentclass{ckm}                 

\usepackage{txfonts}            
\usepackage{multirow}
\input symbols
\newcommand{\thetatr}{\ensuremath{\theta_{\scriptscriptstyle tr}}\xspace}
\newcommand{\RT}{\ensuremath{R_{\scriptscriptstyle T}}\xspace}
\newcommand{\Aperp}{\ensuremath{A_{\perp}}\xspace}
\newcommand{\Apar}{\ensuremath{A_{\parallel}}\xspace}
\newcommand{\Az}{\ensuremath{A_{0}}\xspace}
\def\ctwob{\ensuremath{\cos\! 2 \beta   }\xspace}
\def\DG{\ensuremath{{\rm \Delta}\Gamma}\xspace}
\def\Bd      {\ensuremath{B_d}\xspace}
\def\bphiks  {\ensuremath{\Bz \to \phi \KS}\xspace}
\def\betaks  {\ensuremath{\Bz \to \eta' \KS}\xspace}
\def\bpsikstz    {\ensuremath{\Bz \to \jpsi \Kstarz}\xspace}
\def\bpsipiz     {\ensuremath{\Bz \to \jpsi \piz}\xspace}
\def\bpsiksl     {\ensuremath{\Bz \to \jpsi K^0_{\scriptscriptstyle S,L}}\xspace}
\def\ea     {{\em et al.}\xspace}
\newcommand{\jprb}      [1]  {\jprBase\ B~{\bf #1}}

\confname{Workshop on the CKM Unitarity Triangle, IPPP Durham, April
  2003}

\title{$\sin(2\beta)$: Status and Prospects}

\author{G.~Raven\thanks{supported by F.O.M., program 23 (The Netherlands)}}
\address{NIKHEF and Vrije Universiteit, Amsterdam}

\begin{document}

\begin{abstract}An overview of the observation of \CP violation
in the neutral $B$ system, and the
measurements of the \CP-violating asymmetry \stwob with $B\to$  
charmonium $K^0_{S,L}$ events, 
performed by the \babar\ and Belle experiments at 
the SLAC and KEK $B$ factories is given.
In addition, the measurements of \stwob with several other modes are described,
including $B\to\phi\KS$, which, as the leading contribution is from a loop diagram,
could be sensitive to physics beyond the Standard Model.
\end{abstract}

\maketitle


\section{Introduction}

\CP violation has been a central concern of particle physics since its
discovery in 1964 in the decays of \KL\ decays~\cite{EpsilonK}.
An elegant explanation of the
\CP-violating effects in these decays is provided by
the \CP-violating phase of the three-generation Cabibbo-Kobayashi-Maskawa
(CKM) quark-mixing matrix~\cite{CKM}.
However, existing studies of \CP violation
in neutral kaon decays and the resulting experimental constraints on the
parameters of the CKM matrix~\cite{MSConstraints} do not provide a stringent
test of whether the CKM phase describes \CP violation~\cite{Primer}.
In the CKM picture, large \CP-violating asymmetries are expected to occur in
the time distributions of \Bz\ decays to charmonium final states. 

In general, \CP-violating asymmetries are due to the interference
between amplitudes with a weak phase difference. 
For example,
a state initially produced as a \Bz\ (\Bzb) can decay to a \CP eigenstate,
such as $\jpsi \KS$, either directly,
or it can first oscillate into a \Bzb\ (\Bz) and then decay
to $\jpsi \KS$.  With little theoretical uncertainty in the Standard Model, the 
phase difference between these two amplitudes is equal to twice the angle
$\beta = \arg \left[\, -V_{\rm cd}^{ }V_{\rm cb}^* / V_{\rm td}^{ }V_{\rm tb}^*\, \right]$ 
of the Unitarity Triangle~\cite{bigisanda}.  The measurement of the 
\CP-violating asymmetry in this decay allows a direct determination 
of \stwob, and can thus provide a crucial test of the Standard Model.

Initial measurements of the \CP asymmetry in \bpsiks\ were performed
at LEP by Aleph and Opal, and at the Tevatron by CDF~\cite{stwob_other}, but
the small branching ratio of this decay made it difficult for the
the experiments to obtain sufficient events for a statistically significant
measurement. The KEK and SLAC based $B$ factories, running at the 
\FourS\ resonance, were designed to provide the required high luminosity
to perform this measurement.
Although the measurements from the \babar\ and Belle experiments,
at SLAC respectively KEK, after the first year of running, shown 
in summer of 2000, were not yet conclusive, only a year later both 
experiments were able to claim the observation of \CP violation in 
the $B$ meson system. 
And in 2002 the direct measurements~\cite{stwob} of \stwob\ surpassed
the precision of the indirect determination of $\beta$ obtained from
\CP-conserving variables, assuming the validity of the CKM 
description~\cite{CKMfitter}. 
The consistency of these measurements with their prediction~\cite{parodi}
implies that the CKM description of the \CP violation in the quark sector 
has successfully passed its first quantitative test.

\section{Measurement of \stwob at \FourS\ \B-factory experiments}

A \BzBzb\ pair produced in \FourS\ decays
evolves as a coherent $P$-wave until one of the \B\ mesons decays.
If one of the \B\ mesons, referred to as \Btag, can be 
ascertained to decay to a state of known flavour, {\em i.e.} \Bz\ or
\Bzb, at a certain time $t_{\rm tag}$, 
the other \B, referred to as \Brec, {\it at that time} must be of the
opposite flavour as a consequence of Bose symmetry.
Consequently, the oscillatory probabilities for observing
\BzBzb, $\Bz\Bz$ and $\Bzb\Bzb$ pairs produced in
\FourS\ decays are a function of
$\deltat = t_{\rm rec} - t_{\rm tag}$, allowing
the mixing frequency and \CP asymmetries to be determined
if \deltat\ is known.

The proper-time distribution of $B$ meson decays to a \CP eigenstate
with a \Bz\ or \Bzb\ tag
can be expressed in terms of a complex
parameter $\lambda$ that depends on the both the \BzBzb\ oscillation amplitude
and the amplitudes describing \Bz\ and \Bzb\ decays to this final state.
The decay rate $f_+(f_-)$ when the tagging meson is a \Bz(\Bzb) is given
by 
$$
f_\pm(\deltat) = \frac{e^{-|\deltat|/\tau_{\Bz}} }{4 \tau_{\Bz}} 
                               \left[ 1 \pm S \sin(\deltamd\deltat)
                                        \mp C \cos(\deltamd\deltat)
                               \right],
$$
where $\deltat = t_{\mathrm{rec}} - t_{\mathrm{tag}}$ is the difference
in proper decay times of the reconstructed $B$ meson ($B_{\mathrm{rec}}$)
and the tagging $B$ meson ($B_{\mathrm{tag}}$), $\tau_{\Bz}$ is the \Bz\ lifetime,
and \deltamd is the \BzBzb\ oscillation frequency.
The sine coefficient, which is given by
$S = 2\Im{\lambda}/(1+|\lambda|^2)$ is due to the interference
between direct decay and decay after flavour change, and the cosine
coefficient, $C = (1-|\lambda|^2)/(1+|\lambda|^2)$ is due to the interference
between decay amplitudes with different strong and weak phases.
In the Standard Model, $\lambda_f = \eta_f e^{-2i\beta}$ for charmonium-containing
$ b\to c (\cbar s)$ decays, where $\eta_f$ is the \CP eigenvalue of the final
state $f$.

At asymmetric $e^+e^-$ colliders such as PEP-II at SLAC and KEK-B
at KEK~\cite{accelerators},
resonant production of the \FourS\  provides a copious source of $\BzBzb$ 
pairs moving along the beam axis ($z$ direction)
with an average Lorentz boost $\left<\beta\gamma\right> $ of 0.56 and 0.43
respectively.
Therefore, the proper decay-time difference $\deltat$ is,
to an excellent approximation, proportional to
the distance \deltaz\ between  the two \Bz-decay vertices
along the axis of the boost,
$\deltat \approx \deltaz / c\left< \beta \gamma \right>$.

The analysis of the data proceeds in the following steps:
\begin{enumerate}
\item selection of events where one $B$, referred to as $B_{\mathrm{rec}}$
      is fully reconstructed;
\item determination of the vertex of the other $B$ decay, $B_{\mathrm{tag}}$,
      and computation of \deltat;
\item determination of the flavour of $B_{\mathrm{tag}}$ from its charged
      decay products.
\end{enumerate}
Both experiments determine their \deltat resolution and the mistag rate of 
the flavour tagging algorithms from control samples, obtained from the data
itself.

\subsection{Data samples and \B reconstruction}

\begin{figure*}
\hbox to\hsize{\hss
\includegraphics[width=0.40\hsize]{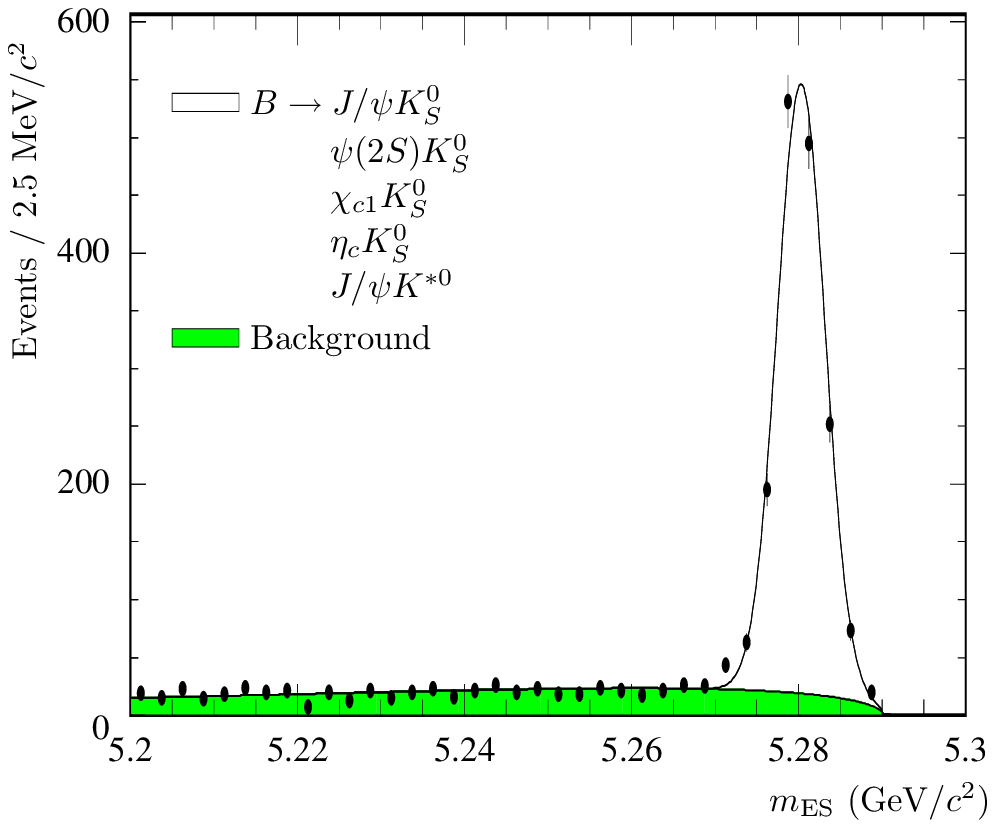}
\includegraphics[width=0.38\hsize]{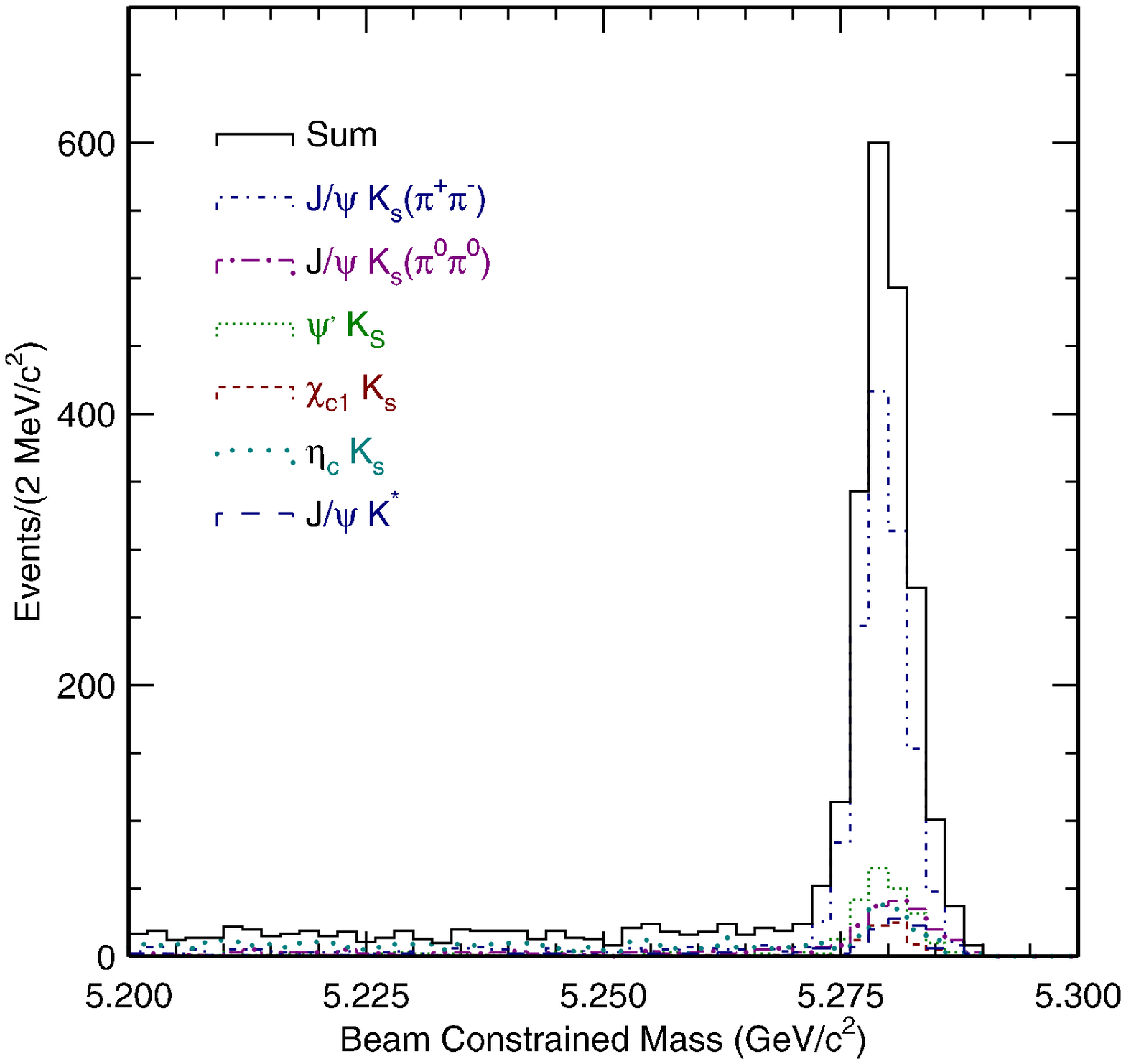}
\put(-220,130){ a) }
\put(-40,130){ b) }
\hss}
\hbox to\hsize{\hss
\includegraphics[width=0.37\hsize]{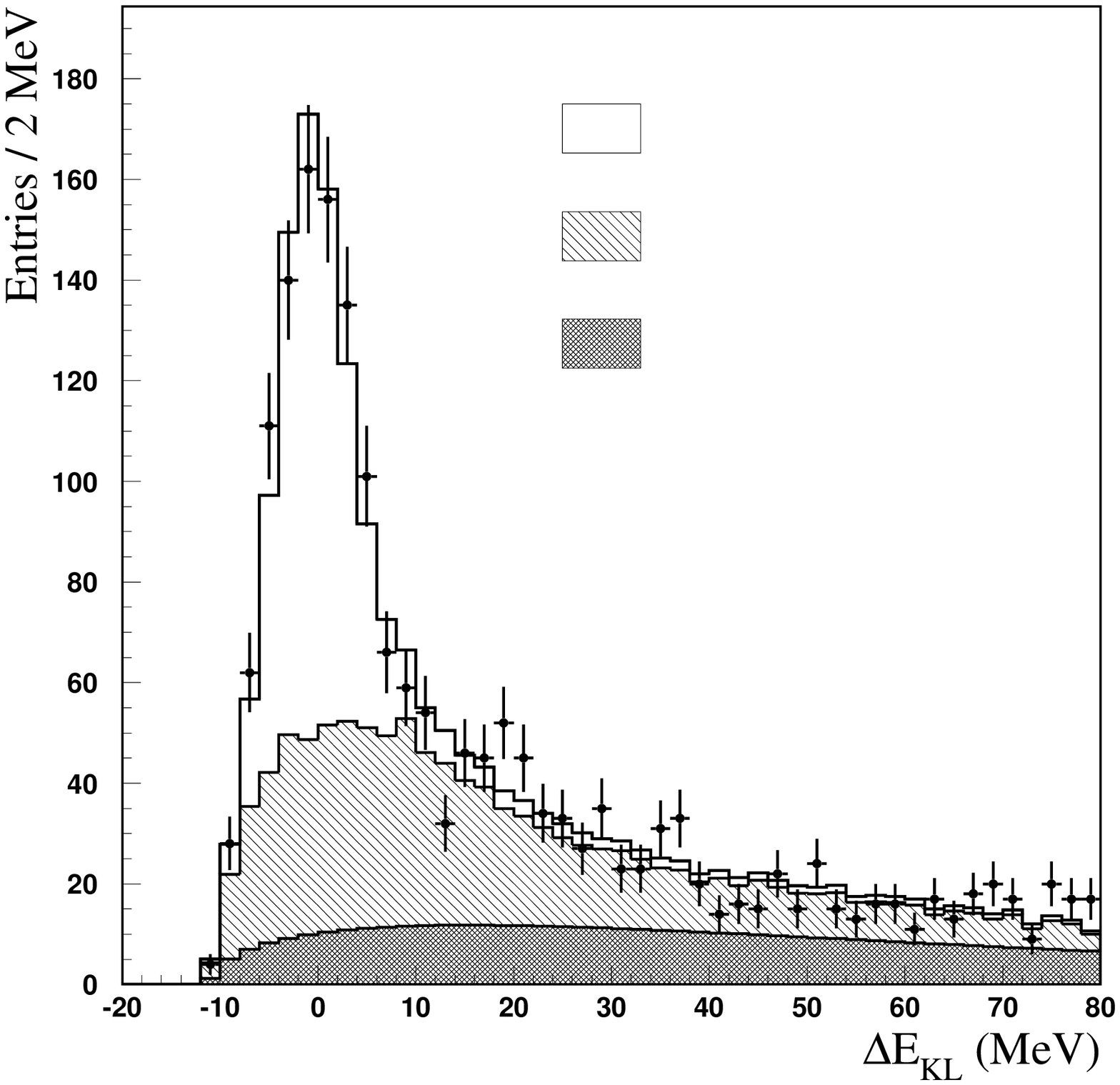}
\includegraphics[width=0.37\hsize]{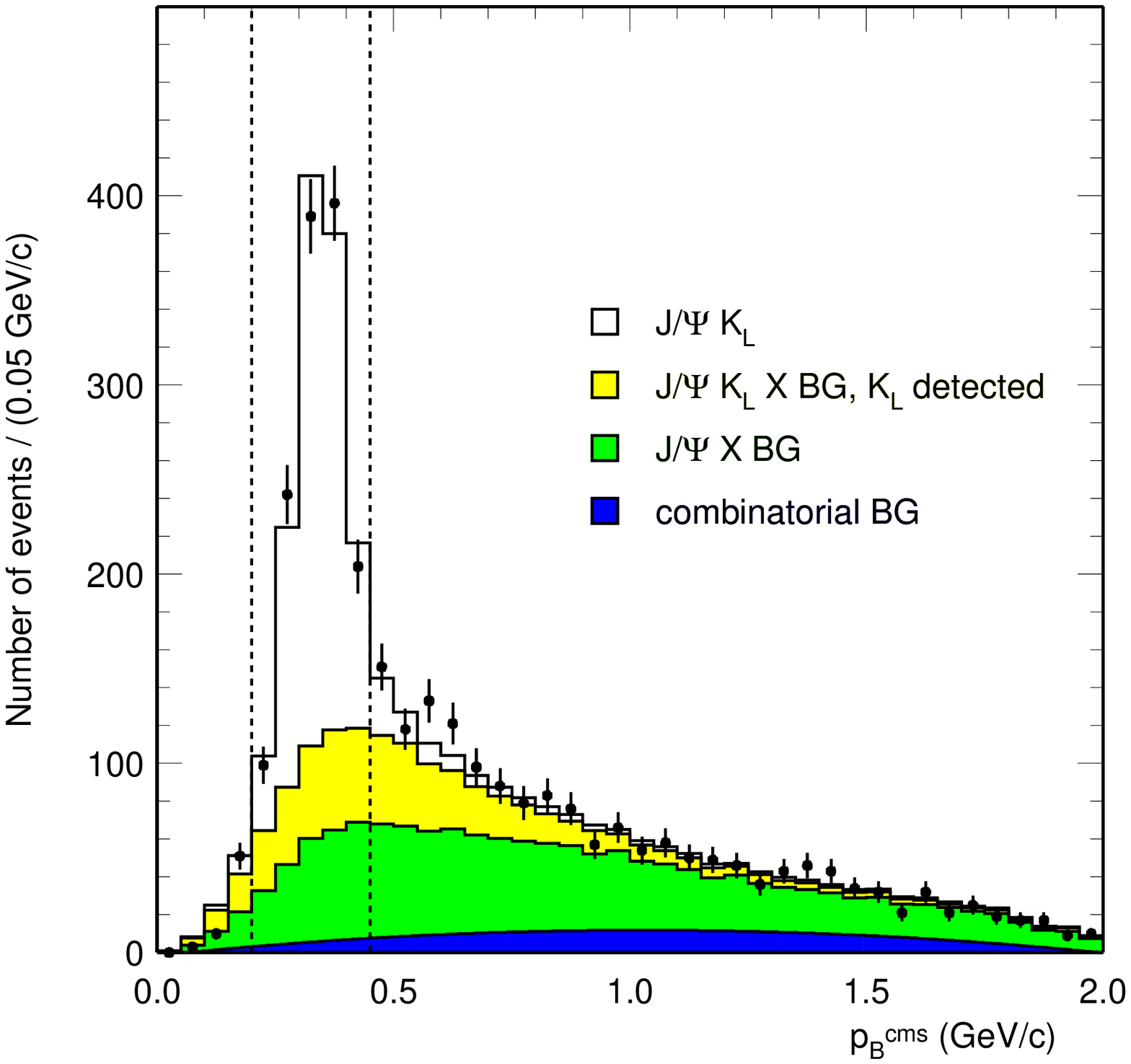}
\put(-220,150){ c) }
\put(-267,142){\tiny \jpsi\KL signal }
\put(-267,125){\tiny \jpsi X background }
\put(-267,110){\tiny non-\jpsi background }
\put(-25,150){ d) }
\hss}
\caption{Distributions of beam-energy substituted mass for charmonium \KS
events, for \babar\ (a) and Belle (b), and \DeltaE (c) and $p^{*}$(d) for 
\jpsi\KL events, for \babar\ and Belle, respectively.}
\label{fig:psiksl_events}
\end{figure*}

Both experiments have sofar published their \stwob\ measurements
on samples obtained as of July 2002. In the case of \babar, this
implies a sample of $88\cdot 10^6$ \FourS decays, whereas Belle 
collected a sample of $85 \cdot 10^6$ decays.
As the branching ratios of decays of $B$ mesons to \CP eigenstates
are small, e.g. a few times $10^{-4}$ for $\jpsi\KS$, 
both experiments increase the size of the event sample by reconstructing
several final states:
$\jpsi\KS,\psitwos\KS,\chicone\KS,\etac\KS,\jpsi\Kstarz(\KS\piz)$
and $\jpsi\KL$. In addition, to determine the performance of the
\deltat\ reconstruction and the flavour tagging, control samples of
fully reconstructed decays of $B$ mesons to self-tagging flavour eigenstates 
are selected\footnote{Throughout this paper, charge-conjugate modes are implied.}:
$\Bz\to D^{(*)-}\pip, D^{(*)-}\rhop, D^{(*)-}a_1^+$ and $\jpsi\Kstarz(\Kp\pim)$.
In addition, semileptonic decays into $\D^{*-}\ell^+\nu$ are selected.
The main selection criteria of the fully reconstructed decays are the
energy difference, \DeltaE, between the energy of 
the reconstructed candidate and the beam-energy in the \FourS\ center-of-mass
system, and the beam-energy substituted mass, \mes, also known as the
beam-constrained mass, defined as $\mes = \sqrt{s/4 - {p^*}^2}$, where
$s$ is the square of the center-of-mass energy and $p^*$ is the momentum of
the $B$ candidate in the center-of-mass. In the case of signal events, these
variables are distributed according to Gaussian distributions, centered 
at $\DeltaE = 0$ and $\mes = m_B$ respectively.
The distributions of $\mes$ for charmonium \KS events are shown in 
Figure~\ref{fig:psiksl_events}. In the case of \jpsi\KL, only the direction
of the \KL\ is measured, and, to determine its momentum, both experiments
constrain the mass of the candidate to the $B$ mass. Next, they plot either
the $p^*$ of the candidate, or \DeltaE. These distributions are also shown in 
Figure~\ref{fig:psiksl_events}.
\begin{table}
\begin{tabular}{|l|cc|cc|}
\hline
                            & \multicolumn{2}{|c}{\babar} & \multicolumn{2}{|c|}{Belle} \\
Mode                        &  $N_{\mathrm{sig}}$ & ${\cal P}$(\%)& $N_{\mathrm{sig}}$  &   ${\cal P}$(\%) \\
\hline
$\jpsi\KS(\pip\pim)       $ & 1429 & 96 &     1116 & 96    \\
other ($\cbar c$)\KS         &  721 & 85 &      523 & 86    \\
$\jpsi\Kstarz(\KS\piz)    $ &  283 & 73 &       89 & 84   \\
\hline
\parbox[l]{1.8cm}{flavour\\ eigenstates}          & 32700& 83 &     18045 & 82 \\
\hline
\end{tabular}
\caption{Number of selected events in the signal region ($N_{\mathrm{sig}}$) and the corresponding
purities (${\cal P}$).}
\end{table}

\subsection{Determination of \deltat}

 The time difference \deltat\ can be related to the
 distance $\Delta z$ along the boost axis between the
 decay points of the two $B$ mesons. Approximating the unmeasured
 sum of the proper times by the average \Bz lifetime, $\tau_B$, yields 
  $\Delta z = \beta\gamma\gamma^*_{\mathrm{rec}}
 c \deltat + \gamma \beta^*_{\mathrm{rec}}\gamma^*_{\mathrm{rec}}
 \cos \theta^*_{\mathrm{rec}} c (\tau_B+|\deltat|)$,
 where $\theta^*_{\mathrm{rec}}$, $\beta^*_{\mathrm{rec}}$ and $\gamma^*_{\mathrm{rec}}$
 are the polar angle with respect to the boost direction,
 the velocity and the boost of the reconstructed $B$ candidate in
 the \FourS\ frame.
  Whereas \babar\ solves the above equation for
 \deltat, Belle makes an approximation which only keeps the
 first term: $\deltat= \Delta z/ c\beta\gamma\gamma^*_{\mathrm{rec}}$.

 As one of the $B$ mesons, $B_{\mathrm{rec}}$ is fully reconstructed, its 
 decay vertex position is well known. The decay vertex of the other $B$ 
 meson, $B_{\mathrm{tag}}$, is inferred from the charged particle tracks 
 remaining after the decay products of $B_{\mathrm{rec}}$ are removed.
 To remove tracks from secondary decays, both experiments first remove 
 tracks from $\KS$ and $\Lambda$ candidates as well as photon conversion,
 and then perform an iterative fit procedure, rejecting those tracks with the large
 contribution to the $\chi^2$. In the case of Belle, the constraint that
 the vertex of $B_{\mathrm{tag}}$ is consistent with the beamspot is applied. 
 \babar\ instead requires that the $B_{\mathrm{tag}}$ vertex is consistent
 with the line of flight computed from the location of the beamspot, 
 the momentum of $B_{\mathrm{rec}}$ and the known \FourS boost.
 The resolution obtained on $\deltat$, determined from the fully reconstructed
 flavour samples, is 1.1 ps for \babar\, and 1.4 ps for Belle, partly due
 to the difference in the \FourS boost.

\subsection{Flavour tagging}

After the daughter tracks of the $B_{\mathrm{rec}}$ are removed from the event,
the remaining tracks are analyzed to determine the flavour of the $B_{\mathrm{tag}}$, and
this ensemble is assigned a flavour tag, either \Bz or \Bzb. For this purpose,
flavour tagging information carried by primary leptons from semileptonic $B$
decays, charged kaons, soft pions from \Dstar\ decays, and more generally by
high momentum charged particles is used.

Belle uses the likelihood ratios of the properties of these particles to estimate
the mistag rate for each individual event, and then ranks events into six mutually
exclusive groups based on their estimated mistag rate. \babar\ uses neural networks,
trained according to each of the physics processes mentioned above, and classifies
events into four mutually exclusive categories according to the underlying physics 
process, combined with performance criteria based on the neural network output.

As the amplitude of the observed \CP asymmetries will be reduced by a
factor $1-2w$, where $w$ is the mistag rate, it is crucial for the experiments
to determine the mistag rates of the various tagging categories from data.
This can be done by considering 
decays to flavour eigenstates, where the deviation of the observed mixing
asymmetry from unity is also given by $1-2w$. \babar\ uses fully 
reconstructed events in the modes $D^{(*)-}h^+ (h^+=\pip,\rho^+,a_1^+)$ and
$\jpsi\Kstarz (\Kstarz\to K^+\pim)$, whereas Belle uses fully 
reconstructed events in modes $D^{(*)-}\pip$ and $D^{*-}\rho^+$, complemented by 
$\Bz\to{\Dstar}^-\ell^+{\nu}_\ell$ events. In the case of Belle, the mistag rates are
determined by fitting the control samples separately, and then propagating
the obtained values to the fit on the \CP sample. The statistical uncertainty
on the mistag rates due to the finite size of the control samples is accounted
for in systematic errors. \babar\ proceeds differently, performing a simultaneous
fit to both the control samples and the \CP sample. This automatically insures
that the statistical error on the mistag rates is propagated into the statistical
error on the \CP asymmetries. Even though the flavour tagging algorithms are
somewhat different between the experiments, their performance is very similar:
the total effective tagging efficiency $Q$, which is given by $Q = \sum_i \eps_i ( 1 - 2 w_i)^2$,
is measured to be $28.6\pm0.6 \%$ for Belle, and $28.1\pm 0.6\%$ for \babar.

One complication has recently received attention, partly
due to its relation to the measurement of $\sin(2\beta+\gamma)$: 
when decays of the type $\B\to\D X$ are used to infer the flavour of
the parent \B mesons, one suffers from an intrinsic mistag rate due to
the contribution of CKM suppressed $b\to u (\cbar d)$ decays.
This effect is put to good use in the measurement of $\sin(2\beta+\gamma)$,
as the suppressed mode can, once \BzBzb oscillations are taken  into account,
interfere with the favoured $\bbar\to \cbar (u \dbar)$ amplitude. 
As the relative weak phase between these decay amplitudes is given by $\gamma$,
the results is a time-dependent \CP asymmetry, depending on 
$\sin(2\beta+\gamma)$, albeit with a magnitude which is suppressed by 
$|V_{ub}^*V_{cd}/V_{cb}^*V_{ud}|^2\approx (0.02)^2$.
This same interference, when applied to the tagging decay 
effectively results in a mistag rate which is {\em not} constant as 
a function of $\deltat$, and thus is not accounted for in the 
experimental determined mistag rate which is assumed to 
be independent of $\deltat$.
However, because the two \B mesons produced
by \FourS decays are correlated until one of them decays, $B_{rec}-B_{tag}$
interference terms involving favoured and suppressed amplitudes
are only suppressed by a factor of about $0.02$.  The result is that for
$|\lambda_{\jpsi\KS}|=1$ the $S_{\jpsi\KS}$ 
and $C_{\jpsi\KS}$ coefficients are now given by~\cite{LongBaak}:
\begin{eqnarray*}
C_{\jpsi\KS} &=& -2 r' \sin\gamma \sin \delta'  \nonumber \\
\label{eq:DCSD}
S_{\jpsi\KS} &=&  \stwob 
   \Big[ 1 -2 r' \cos \delta' 
            \big( 
                 \ctwob\cos(2\beta+\gamma)   \\
           &&{}\;\;\;\;\;\;\;\;\;\;\;\;\;\;\;\;\;\;\;\;\;\;\;\;\;\;\;\;\;+ \kappa\stwob\sin(2\beta+\gamma) 
            \big)
    \Big] \nonumber
\end{eqnarray*}
where  $\delta'$ and $r'$ are the effective strong phase and ratio of the suppressed 
to favored amplitudes obtained when all final states contributing
to a particular tagging category are combined, and $\kappa$, an empirical constant
which depends on the values of $\beta$ and $\gamma$, is approximately 0.3.  
Fortunately, lepton tags are unaffected by this effect, and, as lepton tags
represent about $1/3$ of the effective tagging efficiency, this effect
is suppressed by a factor of $2/3$.  As can be seen from equations above
the largest effect is present
for $C_{\jpsi\KS}$, whereas the extraction of \stwob from $S_{\jpsi\KS}$ is 
not very much affected.
However, 
this effect currently dominates the 
systematic uncertainty on the extraction of $|\lambda_{\jpsi\KS}|$ from $C_{\jpsi\KS}$.
\begin{figure*}[tbh]
\hbox to\hsize{\hss
\includegraphics[width=0.51\hsize]{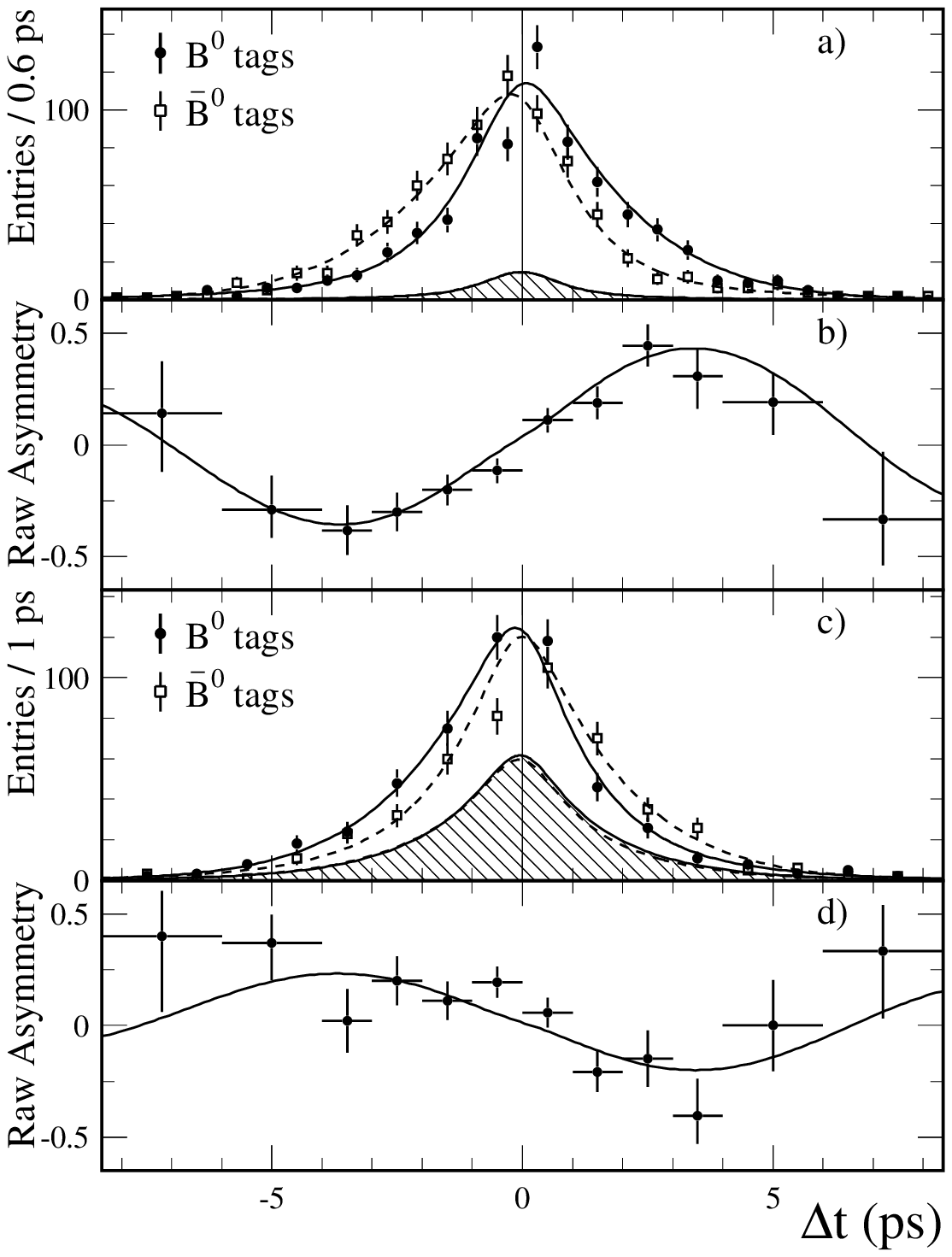}
\includegraphics[width=0.45\hsize]{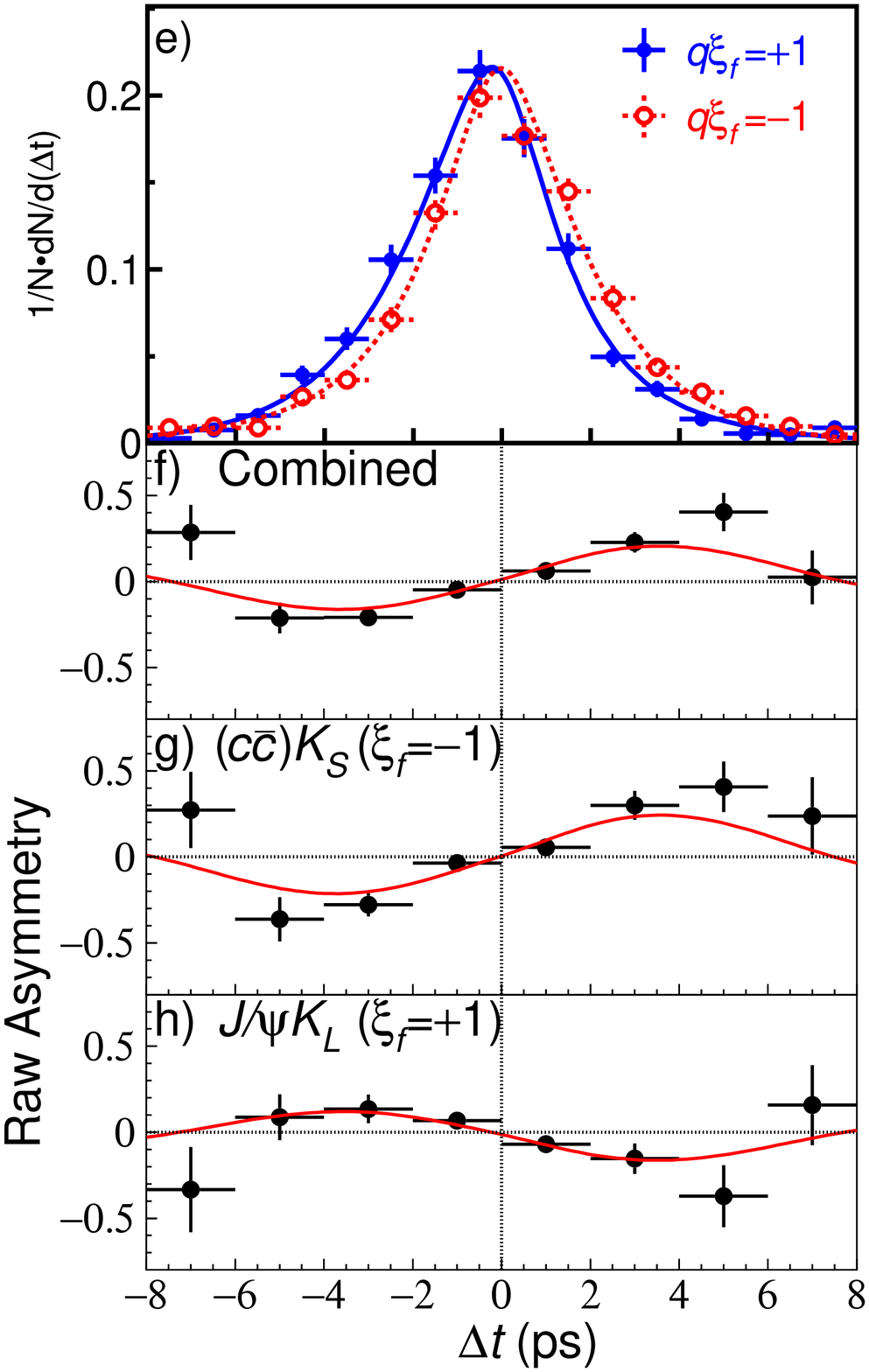}
\hss}
\caption{ The observed \deltat\ distributions for \babar, for
charmonium \KS events (a), and charmonium \KL events(c), and
Belle, for both \KS and \KL combined (e). In addition, the 
asymmetries for charmonium \KS are shown for \babar (b) and Belle (g),
and charmonium \KL, (d) and (h) respectively, and combined (f), 
for Belle.
}
\label{fig:psiksl_asym}
\end{figure*}

\subsection{Current measurements with $b\to c (\cbar s)$ transitions}

\begin{figure}
\hbox to\hsize{\hss
\includegraphics[width=\hsize]{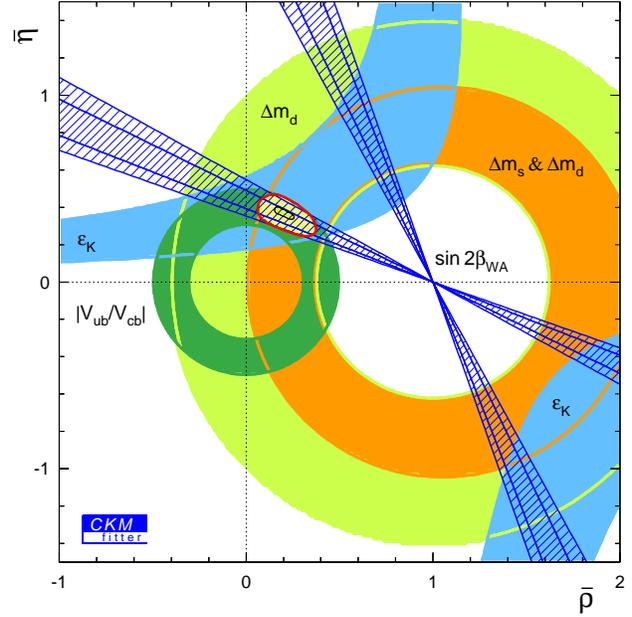}
\hss}
\caption{Constraints on the position of the apex of the Unitarity
Triangle in the $(\bar \rho,\bar\eta)$-plane, including the direct
measurement of \stwob.}
\label{fig:rhoeta}
\end{figure}

    The value of \stwob\ is determined from unbinned maximum-likelihood
fits to the \deltat\ distributions, taking into account the \deltat\
resolution and the mistag rates. 
The projections of the likelihood fits onto the observed \deltat\ distributions
is shown in Figure~\ref{fig:psiksl_asym}. A clear difference in the
\deltat\ distributions for \Bz\ and \Bzb\ tagged events is visible.
The values measured by the two experiments are
\begin{eqnarray*}
    \stwob &=& 0.741 \pm 0.067\;\pm 0.034{\;\;\; \mathrm{ (\babar)} },\\
    \stwob &=& 0.719 \pm 0.074\;\pm 0.035{\;\;\; \mathrm{ (Belle)} },
\end{eqnarray*}
in good agreement with each other. Combining the two measurements yields
\begin{eqnarray*}
\stwob &=& 0.734 \pm 0.055. \nonumber\\
\end{eqnarray*}
The constraint of this measurement on the parameters of the CKM matrix
can be visualized in the $(\bar \rho,\bar \eta)$ plane, as shown 
in Figure~\ref{fig:rhoeta}. 
In addition the constraints derived from \CP-conserving 
measurements and the observed \CP violation in the neutral 
kaon system are included~\cite{CKMfitter}.

\subsection{Extrapolation to larger samples}

Both $B$-factories are performing above expectations, having
accumulated well over 100\invfb each in their first four years of
operation. Currently, PEP-II is capable of routinely delivering 
more than 300\invpb per day, whereas KEK-B has recently set a 
record for daily integrated luminosity of 500\invpb. 
As a result, both experiments are well on their way to collecting 
on the order of 500 \invfb by 2006.
Looking into the past, comparing how the statistical
error on \stwob\ has improved versus the integrated luminosity,
both experiments have been able to perform better
than $\sigma_{\mathrm{stat}}^{-2} \propto \int dt {\cal L}$
by improving their reconstruction, calibrations and selections.
It is however clear that the impact of future improvements, 
other than increased sample size, on the statistical error will 
be less and less pronounced. As a result one can expect a statistical
error on \stwob of approximately $\pm 0.03$ given a 500 \invfb sample.
The main effort will have to be focused on reducing the systematic error.
Currently the
measurement of \stwob\ is still dominated by the statistical
error, but the current systematic uncertainty, even though it is
partly driven by the available sample size, will reach parity 
with the statistical error at the level of about 500 \invfb. 
It is expected that with a combination of additional 
improvements to selections, vertexing and tagging, and further
studies of the data with improved control samples, the systematic 
error can be reduced sufficiently such that the measurement
on 500 \invfb will still be limited by the statistical accuracy.

\section{Approximations in the determination of \stwob}

In the determination of \stwob described above some very reasonable
assumptions are made about both \BzBzb\ mixing and the decay amplitudes
\bpsiks and \bpsikl.
The evolution of the \Bz\ and \Bzb\ states prior to their decay 
is described by oscillations \Bz\to\Bzb and \Bzb\to\Bz with a 
frequency given by the mass difference $\deltamd = m_H - m_L$ of the 
\Bd mass eigenstates, multiplied by factors $q/p$ and $p/q$, respectively.
In the measurement of \stwob, it is assumed that $|q/p|=1$, which, given
the Standard Model expectation of $|q/p|-1 = (2.5 - 6.5) \times 10^{-4}$
for the \Bd\ system~\cite{qoverp}, is  a very good approximation. 
If $|q/p|=1$, the rate of \Bz\to\Bzb and \Bzb\to\Bz should be equal,
unlike the case for the neutral Kaon system.  This possible rate difference
can be determined by measuring the like-sign lepton asymmetry, 
${\cal A}_{sl}=\left(N_{\ell^+\ell^+}-N_{\ell^-\ell^-}\right)/\left(N_{\ell^+\ell^+}+N_{\ell^-\ell^-}\right) 
= \left(1-\left|q/p\right|^4\right)/\left(1+|q/p|^4\right)$.
Several measurements of this asymmetry are available~\cite{other_dilepton}, and
recently this asymmetry has also been measured by \babar~\cite{babar_dilepton} 
to be ${\cal A}_{sl}=(0.5 \pm 1.2  \pm 1.4) \times 10^{-2} $, which 
corresponds to  $|q/p| = 0.998 \pm 0.006  \pm 0.007$.

Recently \babar\ has also determined $|q/p|$ using samples of flavour tagged,
fully reconstructed decays of \Bd\ mesons to either \CP or flavour 
eigenstates~\cite{babar_Fernando}.  Although the sensitivity to $|q/p|$ is
less than for a like-sign dilepton analysis, these samples allow one to also
set a limit on the lifetime difference \DG between the mass eigenstates
and on the complex CPT violating parameter $z$, which is proportional to 
the mass- and lifetime differences between \Bz\ and \Bzb states. 
In the Standard Model, CPT is conserved, and $\DG/\Delta m$ is expected 
to be ${\cal O}\left(m_b^2/m_t^2 \right)$~\cite{dighe}, and thus both 
effects are neglected in the extraction of \stwob.
Within the limited uncertainties of this measurement, no deviations from 
the Standard Model expectations of $z$ and \DG are observed.

An additional assumption made in identifying the sine coefficient
of the time-dependent \CP asymmetry in \bpsiks as \stwob is that 
the decay itself is dominated by a single weak phase. This is an 
excellent approximation as the leading penguin contributions have
the same weak phase as the CKM favoured tree diagram. 
This assumption can to some extent be tested by considering the decay
\bupsik, which is related to \bpsiks by exchange of the spectator
quark. In case there would be a sizable contribution from diagrams
with a different weak phase, there might be a non-zero charge 
asymmetry: ${\cal A}_{+-}(\jpsi K^\pm) = \left(N_{\jpsi K^+} - N_{\jpsi K^-}\right)
/\left(N_{\jpsi K^+} - N_{\jpsi K^-}\right).$
This asymmetry has been measured by both \babar\ and Belle~\cite{jpsikp}, and 
the values obtained are consistent with zero:
\begin{eqnarray*}
{\cal A}_{+-}(\jpsi K^\pm) &=& +0.003\pm0.030\pm 0.004{\;\;\; \mathrm{(\babar)}}\\
{\cal A}_{+-}(\jpsi K^\pm) &=& -0.042\pm0.020\pm 0.017{\;\;\; \mathrm{(Belle)}}
\end{eqnarray*}

The decays \bpsiks and \bpsikl both proceed through CKM favoured, 
colour suppressed tree diagrams $\Bz\to \jpsi \Kz$ , followed by $\Kz\to\KS$ and
$\Kz\to\KL$ respectively. As a result, neglecting the tiny amount
of \CP violation in neutral kaon mixing, the time dependent asymmetries
in \bpsiks and \bpsikl should be equal in magnitude, but opposite in
sign, ${\cal A}_{\bpsiks} = -{\cal A}_{\bpsikl}$. It
can be shown that to generate a deviation of more 
than a few times $10^{-3}$, interference between the favoured decay and a so-called 
wrong flavour decay, \Bz\to\jpsi\Kzb, is required~\cite{grkali}. 
By considering the related decay \bpsikstz, with \Kstarz decaying to
\Kp\pim, one can tag the kaon flavour in the decay, and by performing
a time-dependent analysis \babar\ measures the following ratios of 
wrong-flavour to favoured amplitudes~\cite{Gautier}:
\begin{eqnarray*}
\Gamma(\bar\Bz\!\to\!\jpsi    \Kstarz)/ \Gamma(    \Bz\!\to\!\jpsi\bar\Kstarz)\!\!\!\!&=&\!\!\!\!\!\!-0.022\pm0.028\pm 0.016,\\
\Gamma(    \Bz\!\to\!\jpsi\bar\Kstarz)/ \Gamma(\bar\Bz\!\to\!\jpsi\bar\Kstarz)\!\!\!\!&=&\!\!\!\!\!\!\;\;0.017\pm0.026\pm 0.016.
\end{eqnarray*}
Again, no evidence for a deviation from the Standard Model expectations is observed.

\section{Measurement of \ctwob with \bpsikst}

The decay of \bpsikstz,\Kstarz\to\KS\piz proceeds through two \CP-even 
amplitudes (\Az ,\Apar) and one \CP-odd amplitude (\Aperp). 
This implies that, unless one takes into account the angular
dependence of the contributing amplitudes, the magnitude of the
\CP asymmetry is diluted by an additional factor $1-2\RT$, where
$\RT$ is the fraction of \CP-odd decay rate. 
The simplest way to extract \stwob from these decays is to measure
\RT, and insert the additional dilution $1-2\RT$ in the time dependent
analysis. 
Both \babar\ and Belle have measured \RT~\cite{kstarRt}, and
the combined results shows that this decay is mostly \CP-even, $\RT = 0.179 \pm 0.030$.
One can improve the sensitivity by taking into account the dependence
of \CP-even and odd amplitudes on $\cos(\thetatr)$,
where \thetatr is the angle in the \jpsi rest-frame between the positive
lepton and the normal to the decay plane of the \Kstarz:
the \CP-even components are proportional to $1-\cos^2\thetatr$, and 
the \CP-odd component is proportional to $(1+\cos^2\thetatr)/2$.
A further refinement can be obtained by including all three angles
that describe this decay. Denoting the three observable angles 
in this decay by $\vec{\omega}$, the decay rate is
given by~\cite{babar_moriond02,itoh_vv}:
\begin{eqnarray*}
f_\pm(\deltat,\vec{\omega})  &\propto &
 \frac{ e^{-|\deltat|/\tau_{\Bz} }}{4 \tau_{\Bz}} \times   \\ 
 &&\Big[  {\cal I}\left(\vec{\omega},\vec{A})\right) 
       \mp{\cal C}\left(\vec{\omega},\vec{A})\right)\cos\deltamd\deltat \\
   &&  \pm \big\{
   {\cal S}_{\sin}\left(\vec{\omega},\vec{A}\right)\stwob \\
   &&{}\;\;\;\;       + {\cal S}_{\cos}\left(\vec{\omega},\vec{A}\right)\ctwob 
  \big\}\sin\deltamd\deltat\Big]
\end{eqnarray*}
and at first sight one expects to be able to determine 
\ctwob. This would allow one to eliminate two of the four
ambiguities in $\beta$ from the measurement of \stwob. Unfortunately
the observable ${\cal S}_{\cos}\left(\vec{\omega},\vec{A}\right)\ctwob$ is invariant under the transformation
$(\phi_\perp ,\phi_\parallel,\ctwob)\to(\pi-\phi_\perp,-\phi_\parallel,-\ctwob)$,
where $\phi_i$ are the relative phases between $A_i$.
As a result one can only determine the sign of \ctwob if one
could choose between the two possible solutions for the strong phases.
The two experiments quote both ambiguities~\cite{babar_moriond02,itoh_vv}, including
the corresponding strong phases:
\begin{eqnarray*}
 \ctwob\!\!\!\! &=&\!\!\!\!\! \left\{ \!\! \begin{array}{l}
                   +3.3 {}^{+0.6}_{-1.0}\pm 0.7 \; (\phi_\perp = -0.2,\phi_\parallel = +2.5) \\
                   -3.3 {}^{+1.0}_{-0.6}\pm 0.7 \; (\phi_\perp = -3.0,\phi_\parallel = -2.5)  
                 \end{array}\right.\!\!\! (\babar) \\
 \ctwob\!\!\!\! &=&\!\!\!\!\! \left\{ \!\! \begin{array}{l}
                   +1.4 \pm 1.3 \pm 0.2\;  (\phi_\perp =-0.1,\phi_\parallel = +2.8) \\
                   -1.4 \pm 1.3 \pm 0.2\;  (\phi_\perp =-3.1,\phi_\parallel = -2.8)
                 \end{array}\right.\!\!\!{\mathrm{(Belle)}} \\
\end{eqnarray*}
Thus reducing the number of ambiguities in $\beta$ will require additional
information on which strong phase solution to pick. 
For example, assuming \s-quark helicity conservation~\cite{jpsikstarphase}, the
positive solution seems preferred, but even then the current errors on \ctwob
are still too large to rule out negative values.

\section{Modes with penguin contributions}

\subsection{\bpsipiz}

In the case of \bpsipiz, the tree diagram is CKM suppressed compared to
\bpsiksl. One has thus the possibility that this mode receives 
non-negligible contributions from penguin diagrams with a weak phase
different from the tree diagram. Both $B$-factory experiments have 
observed this decay and determined $S_{\jpsi\piz}$ and $C_{\jpsi\piz}$~\cite{jpsipi0}:
\begin{eqnarray*}
    S_{\jpsi\piz} &=& \;\; 0.05 \pm 0.49 \pm 0.16 {\;\;\; \mathrm{ (\babar)} },\\
    S_{\jpsi\piz} &=&     -0.93 \pm 0.49 \pm 0.08 {\;\;\; \mathrm{ (Belle)} },\\
    C_{\jpsi\piz} &=& \;\; 0.38 \pm 0.51 \pm 0.09 {\;\;\; \mathrm{ (\babar)} },\\
    C_{\jpsi\piz} &=& \;\; 0.25 \pm 0.39 \pm 0.06 {\;\;\; \mathrm{ (Belle)} }.
\end{eqnarray*}
The precision is such that more data is needed to draw a conclusion on 
the possible penguin contribution to the \CP asymmetries in this channel.

\subsection{\bphiks}
There is considerable interest in decays where the leading contribution
to the amplitude is due to loop diagrams, as new
physics processes could provide significant contributions. An example 
are transitions of the type the $\b\to \s (\sbar \s)$ and $\b\to \s (\dbar \d)$,
which are given by gluonic penguin decays, and for which the dominant
penguin contribution has the same phase as $b\to c(\cbar s)$.
As a result, the process \bphiks should exhibit the same \CP asymmetry as
\bpsiks. However, even in the Standard Model there are diagrams with 
different weak phases which contribute to the decay \bphiks, 
but one can set limits on their magnitude 
using isospin related decays such as $\Bu\to\phi\pip$ and $\Kstarz\Kp$.
As a result one expects that within the Standard Model the deviation 
of $S_{\phi\KS}$ from \stwob should  be less than 5\%~\cite{grossman_phiks}. 
Again both experiments have observed clear signals in this mode and measured 
the \CP asymmetries~\cite{Gautier,belle_penguinCP}:
\begin{eqnarray*}
    S_{\phi\KS} &=&    -0.18 \pm 0.51 \pm 0.07 {\;\;\; \mathrm{ (\babar)} },\\
    S_{\phi\KS} &=&    -0.73 \pm 0.64 \pm 0.22 {\;\;\; \mathrm{ (Belle)} },\\
    C_{\phi\KS} &=&    -0.80 \pm 0.38 \pm 0.12 {\;\;\; \mathrm{ (\babar)} },\\
    C_{\phi\KS} &=& \;\;0.56 \pm 0.41 \pm 0.16 {\;\;\; \mathrm{ (Belle)} }.
\end{eqnarray*}
In addition, Belle has measured the time-dependent asymmetries for
the non-resonant $K^+K^-\KS$ final state, and obtains
\begin{eqnarray*}
    S_{KK\KS} &=& 0.49 \pm 0.43 \pm 0.11 {}^{+0.33}_{-0.0},\\
    C_{KK\KS} &=& 0.40 \pm 0.33 \pm 0.10 {}^{+0.26}_{-0.0}.
\end{eqnarray*}
Although the measurements show a trend for smaller or even negative values for $S$, the difference
with \stwob is not yet statistically significant.

\subsection{\betaks}
A mode which is similar to \bphiks is \betaks, but with the additional
complication of a contribution of a CKM suppressed tree-level $\b\to\u$
contribution.
Several estimates of the relative magnitude of the penguin
diagram exist, and the deviation of $S_{\eta'\KS}$ from  \stwob
is expected to be less than ${\cal O}(5\%)$~\cite{SMetapks}.
Both experiments observe clear signals for this mode, and measure the 
time-dependent asymmetries~\cite{belle_penguinCP,etapKs}: 
\begin{eqnarray*}
    S_{\eta'\KS} &=& \;\; 0.02 \pm 0.34 \pm 0.03  {\;\;\; \mathrm{(\babar)}},\\
    S_{\eta'\KS} &=&\;\;  0.71 \pm 0.37 {}^{+0.05}_{-0.06}  {\;\;\;\mathrm{(Belle)}},\\
    C_{\eta'\KS} &=&\;\;  0.10 \pm 0.22 \pm 0.03  {\;\;\; \mathrm{(\babar)}},\\
    C_{\eta'\KS} &=&     -0.26 \pm 0.22 \pm 0.04  {\;\;\; \mathrm{(Belle)} }.
\end{eqnarray*}
Again, no statistically significant deviations from \stwob\ respectively zero 
are observed.

\subsection{\Bztodstdst}

The dominant
contribution to this decay is the transition $b\to c (\cbar d)$, but the
presence of penguin contributions could cause deviations of $S_{D^*D^*}$
from \stwob of about 2\%~\cite{PhamXing}.
Similarly to \bpsikstz, the decay \Bztodstdst is a vector-vector decay which
receives contributions from three partial waves, and either an angular analysis
or a measurement of the \CP-odd fraction \RT is required to interpret the \CP asymmetry. 
From the distribution of $\cos\thetatr$, \babar\ determines
$\RT= 0.07\pm0.06\pm 0.03$, and Belle concludes that the decay
is dominantly \CP-even~\cite{itoh_vv,dstdst}. \babar\ proceeds to measure the time-dependent
\CP asymmetry and finds
\begin{eqnarray*}
    S_{D^*D^*} &=&  0.32 \pm 0.43  \pm 0.13,\\
    C_{D^*D^*} &=&  0.02 \pm 0.25  \pm 0.09.
\end{eqnarray*}

\subsection{\Bztodstd}

This decay, like \Bztodstdst, is a $b\to c (\cbar d)$, but in
this case the final state is not a \CP eigenstate. However, it 
is still possible to determine the \CP asymmetries~\cite{gronau}.
\babar\ has measured the following time-dependent asymmetries~\cite{Gautier}:
\begin{eqnarray*}
    S_{D^{*+}D^-} &=& -0.82 \pm 0.75  \pm 0.14,\\
    S_{D^{*-}D^+} &=& -0.24 \pm 0.69  \pm 0.12,\\
    C_{D^{*+}D^-} &=& -0.47 \pm 0.40  \pm 0.12,\\
    C_{D^{*-}D^+} &=& -0.22 \pm 0.37  \pm 0.10.\\
\end{eqnarray*}
In addition, the time-integrated charge asymmetry has been measured
by \babar\ to be ${\cal A} = -0.03\pm0.11\pm 0.05$. Again, no significant
deviation from the Standard Model expectation is observed.

\section{Conclusion}

\begin{figure*}
\hbox to\hsize{\hss
\includegraphics[width=0.48\hsize]{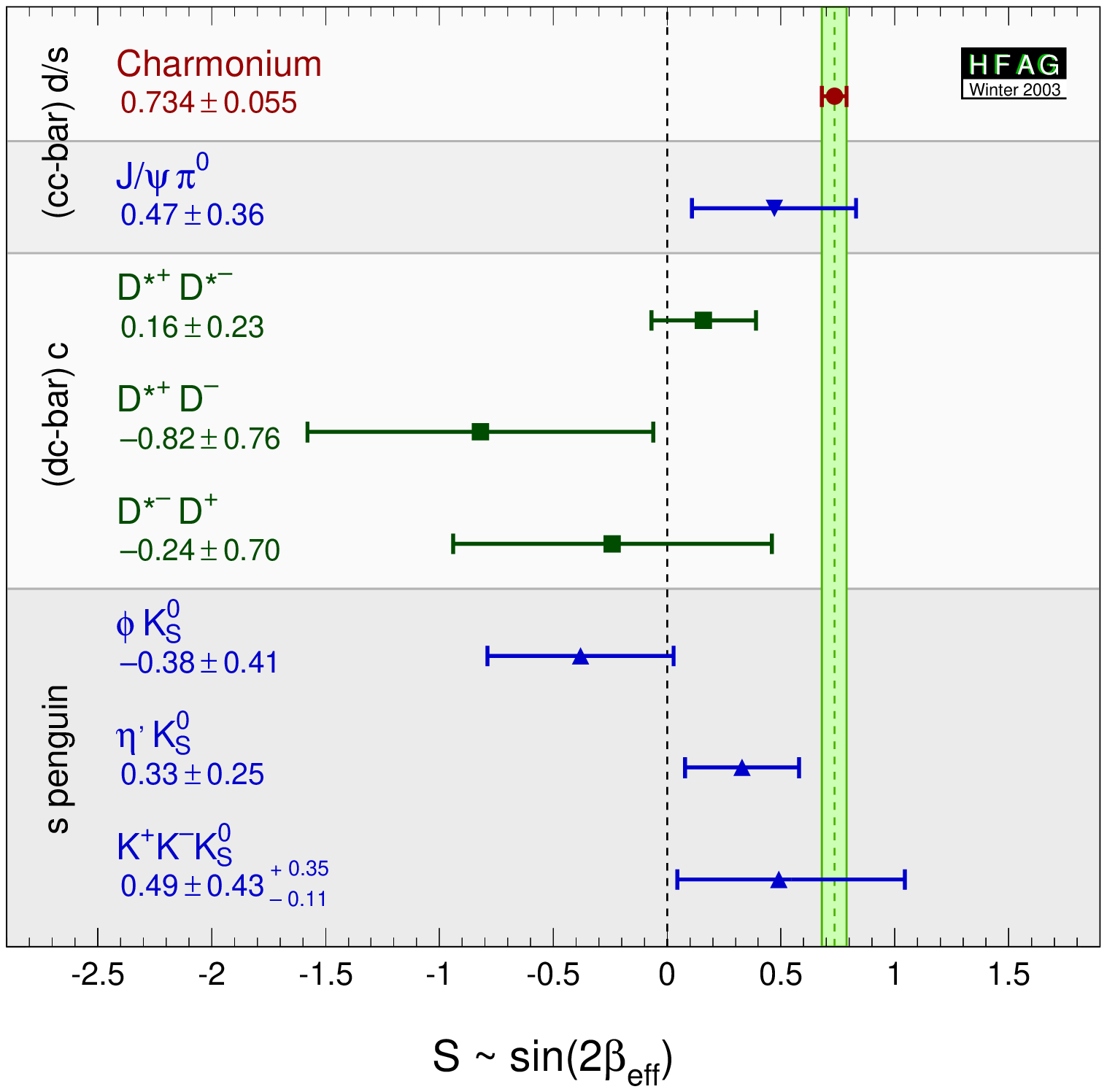}
\includegraphics[width=0.48\hsize]{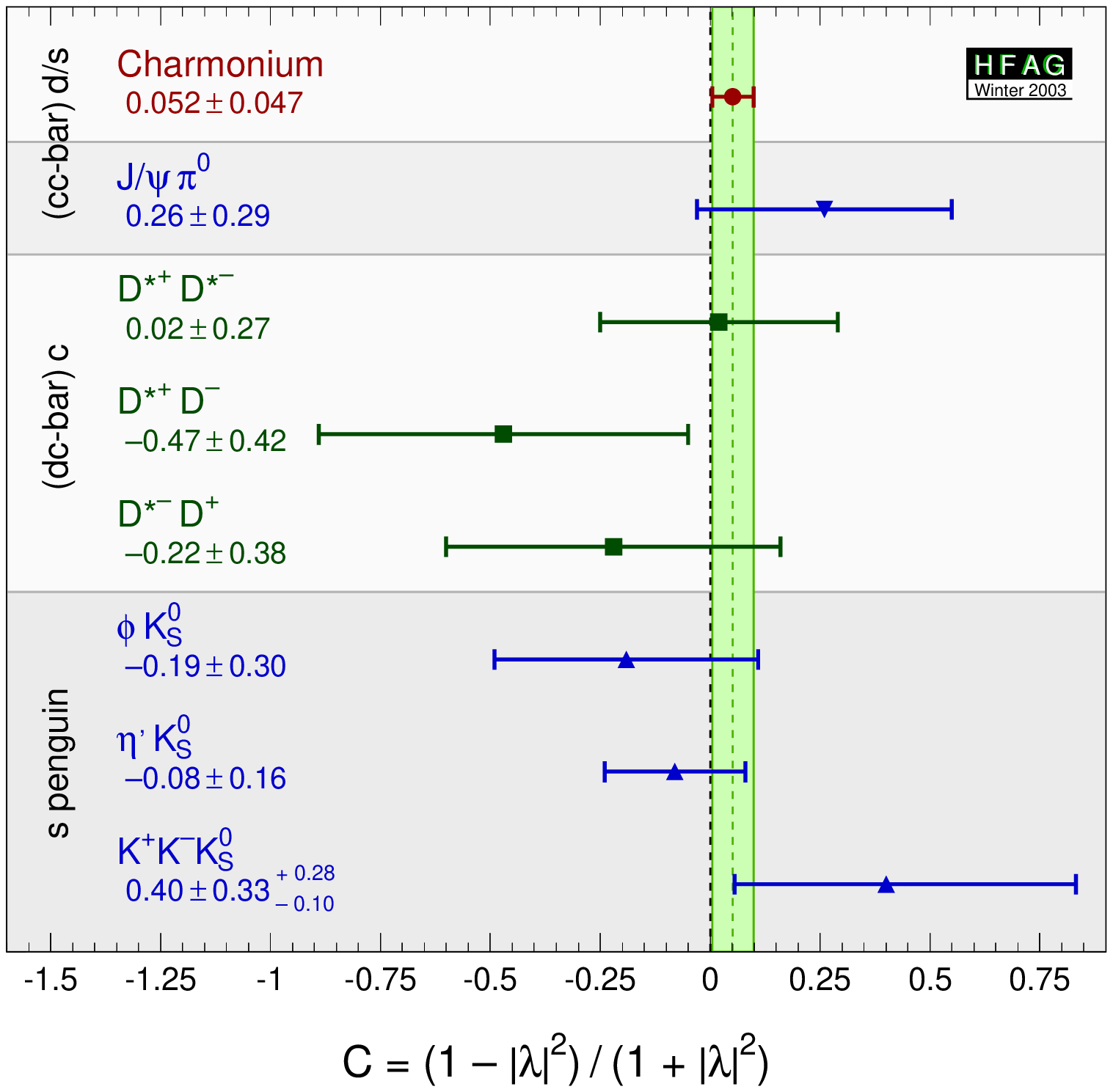}
\hss}
\caption{Summary of the measured $S$ and $C$ coefficients for
the various decay channels, averaged over both \babar\ and Belle~\cite{hfag}.}
\label{fig:SCsummary}
\end{figure*}

The determination of time dependent \CP-violating asymmetries at asymmetric 
energy $B$ factories has reached maturity: the measurement of
\stwob with \bpsiksl\ is dominated by Belle and \babar. 
In a short time we have gone from the first observation of
\CP violation in the B system, to the point where the precision of
the direct measurements of \stwob has exceeded the prediction 
from the indirect measurements.
The $B$ factory experiments have started measuring time-dependent asymmetries
in rare modes such as \bphiks. In the Standard Model, the asymmetries
in these modes are, upto small corrections, equal to \stwob.
A summary of these measurements, averaged over the experiments~\cite{hfag}
is shown in Figure~\ref{fig:SCsummary}.
There is an intriguing trend for these measurements to be lower than
expected, but the current experimental errors are such that no firm 
conclusion can be drawn yet. 
It will be interesting to see whether these measurements will converge, 
as additional luminosity is collected, towards the value of \stwob\ measured with
\bpsiksl, or whether they will become significant deviations, indicating 
the presence of New Physics.

\end{document}